# Sources of Superlinearity in Davenport-Schinzel Sequences

Seth Pettie

The University of Michigan

**Abstract**

A *generalized* Davenport-Schinzel sequence is one over a finite alphabet that contains no subsequences isomorphic to a fixed *forbidden subsequence*. One of the fundamental problems in this area is bounding (asymptotically) the maximum length of such sequences. Following Klazar, let $\text{Ex}(\sigma, n)$ be the maximum length of a sequence over an alphabet of size $n$ avoiding subsequences isomorphic to $\sigma$. It has been proved that for every $\sigma$, $\text{Ex}(\sigma, n)$ is either linear or very close to linear; in particular it is $O(n2^{\alpha(n)^{O(1)}})$, where $\alpha$ is the inverse-Ackermann function and $O(1)$ depends on $\sigma$. However, very little is known about the properties of $\sigma$ that induce superlinearity of $\text{Ex}(\sigma, n)$.

In this paper we exhibit an infinite family of independent superlinear forbidden subsequences. To be specific, we show that there are 17 *prototypical* superlinear forbidden subsequences, some of which can be made arbitrarily long through a simple padding operation. Perhaps the most novel part of our constructions is a new succinct code for representing superlinear forbidden subsequences.

# 1 Introduction

Standard Davenport-Schinzel sequences [10] (or *DS* sequences) are those avoiding a fixed length alternating subsequence of the form $abab\cdots$. The primary applications of these sequences are in bounding the complexity of geometric objects, particularly the lower envelopes of line segments or arbitrary functions with a bounded number of crossings; Agarwal and Sharir [2] have an excellent monograph on geometric applications of DS sequences. It is not difficult to prove that $\text{Ex}(abab, n) = \Theta(n)$, though for longer forbidden subsequences the problem of asymptotically bounding the length of the longest DS sequence is not easy. A celebrated result of Hart and Sharir [12] showed that $\text{Ex}(ababa, n) = \Theta(n\alpha(n))$ where $\alpha$ is the slowly growing inverse of Ackermann's function. It follows that $\text{Ex}((ab)^k, n)$ and $\text{Ex}((ab)^k a, n)$ are also superlinear in $n$ for all $k \geq 3$, though *how* superlinear has still not been completely resolved. Agarwal, Sharir, and Shor [3] gave nearly tight bounds on the length of standard DS sequences:

$$\text{Ex}((ab)^k, n) = n \cdot 2^{\Theta(\alpha(n)^{k-2})}$$

$$\text{Ex}((ab)^k a, n) = \begin{cases} n \cdot \alpha(n)^{O(\alpha(n))^{k-2}} \\ n \cdot 2^{\Omega(\alpha(n))^{k-2}} \end{cases}$$

A natural generalization of DS sequences is to consider *arbitrary* forbidden subsequences, not necessarily those of the form $abab\cdots$.[1] At the moment we have only a limited understanding of how $\text{Ex}(\sigma, n)$ *could* behave, and how it does behave for specific $\sigma$. By generalizing the upper bounds of Agarwal et al. [3], Klazar [16] showed that $\text{Ex}(\sigma, n) = n \cdot 2^{\alpha(n)^{O(1)}}$, where the $O(1)$ depends on $\sigma$. However, there are no commensurate lower bounds, i.e., no specific $\sigma_c \not\succ ababa$ for which $\text{Ex}(\sigma_c, n) \geq n \cdot 2^{(\alpha(n))^c}$. (The notation $x \prec y$ and $x \not\prec y$ mean, respectively, that $x$ is and isn't isomorphic to a subsequence of $y$.) A more basic question—and the subject of this paper—is to identify the features of a forbidden subsequence $\sigma$ that cause $\text{Ex}(\sigma, n)$ to be superlinear. One can see that the set of all superlinear forbidden subsequences can be characterized by a unique set of *minimal* such forbidden subsequences. We define $\Phi$ to be this set:

**Definition 1.1** $\Phi$ *is the smallest set of sequences such that:*

$$\text{Ex}(\sigma, n) = \omega(n) \quad \text{if and only if} \quad \exists \hat{\sigma} \in \Phi : \hat{\sigma} \prec \sigma$$

---

[1] This idea was even suggested by Davenport and Schinzel; see [18] for a discussion of this.

It is amazing how little we know about $\Phi$. Hart and Sharir's result [12] shows that $ababa \in \Phi$ and Adamec et al. [1] showed that no other sequence in $\{a,b\}^*$ is in $\Phi$. Klazar [17] showed that $\Phi$ contains at least two elements: $ababa$ and another which is currently unknown, but is a subsequence of $abcbadadbcd$. In other words, the presence of $ababa \prec \sigma$ is not the *sole* cause of superlinearity in $\text{Ex}(\sigma, n)$. Klazar's result [17] is actually more general in that he shows that any $\sigma$ for which $G(\sigma)$ is strongly connected has $\text{Ex}(\sigma, n) = \omega(n)$, where $G(\sigma)$ is a digraph derived from the syntactic structure of $\sigma$; see Figure 1. In other words, [17] raised the possibility that *strong connectivity* is the sole cause of superlinearity in generalized Davenport-Schinzel sequences.

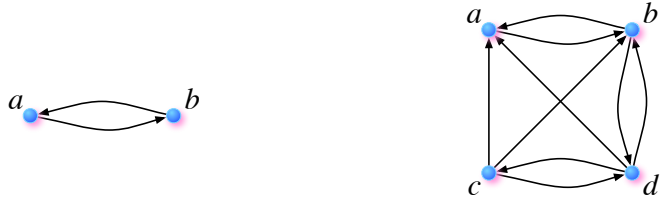

Figure 1: The digraph $G(\sigma)$ has one vertex for each letter in the alphabet of $\sigma$. Assuming that $\sigma$ is repetition-free, an edge $(u,v)$ appears in $G(\sigma)$ if $\sigma$ contains as a subsequence either $vuvu$ or $uvvu$. Left: $G(ababa)$; Right: $G(abcbadadbcd)$.

**New Results.** In this paper we introduce an infinite set $\Psi$ of independent superlinear forbidden subsequences. The elements of $\Psi$ are not fundamentally different but, in fact, naturally divide themselves into just 17 categories. We call the simplest elements of each category the *prototypes*; the elements of $\Psi$ are $ababa$, the prototypes, and an infinite number of sequences that can be derived from the prototypes through a padding operation.

Why 17? There is no particularly good explanation for this number, except that it comes from a new compact notation we use for expressing forbidden subsequences. Whereas a forbidden subsequence is a string over an arbitrarily large alphabet, we can express elements of $\Psi$ as relatively short strings over the fixed alphabet $\{\heartsuit, \spadesuit, \diamondsuit, \clubsuit, \star, (,)\}^*$ that follow some grammatical rules. Grammatical strings correspond to superlinear forbidden subsequences and there just happens to be 17 natural classes of grammatical strings.

Our result refutes the possibility that strong connectivity is *the* cause of superlinearity and addresses an open question posed by Klazar [18], namely, is $\Phi$ infinite or finite? We are able to show that $|\Phi| \geq 5$, though we cannot identify any particular members of $\Phi$ except $ababa$. In Section 6 we discuss why the infinitude of $\Psi$ supports the proposition that $\Phi$ is infinite.

**Related Work.** Davenport-Schinzel sequences are part of a class of problems concerning combinatorial objects with forbidden substructures. Klazar [18] surveys generalizations of DS sequences to trees, permutations, hypergraphs, 0-1 matrices, ordered digraphs, and partitions. Other examples of objects with forbidden substructures are matrices with the Monge property [6] and (partially defined) monotone matrices [4, 15, 14].

Whereas the subject of this paper is finding the causes of superlinearity in DS-sequences, Adamec et al. [1] and Klazar and Valtr [19] looked for specific causes of linearity. In [1] it is shown that $\text{Ex}(abbaab, n) = O(n)$, which implies that $\text{Ex}(a^k b^k a^k b^k, n) = O(n)$ as well, for any $k$. A corollary of this result is that $ababa$ is the only two-symbol sequence in $\Phi$. Klazar and Valtr [19] demonstrated that a few rules (resembling a context free grammar) suffice to generate a huge variety of linear forbidden subsequences. Specifically, let $\bar{\Phi}$ be the set of linear forbidden subsequences, i.e., those that do not contain some element of $\Phi$ as a subsequence. Klazar and Valtr showed that:

$$\begin{aligned}
a^k &\in \bar{\Phi} & &a \text{ any symbol} \\
\sigma_1 a^2 \sigma_2 a \in \bar{\Phi} &\Rightarrow \sigma_1 ab^k a\sigma_2 ab^k \in \bar{\Phi} & &b \text{ a symbol not appearing in } \sigma_1 a^2 \sigma_2 a \\
\hat{\sigma}, \sigma_1 a^2 \sigma_2 \in \bar{\Phi} &\Rightarrow \sigma_1 a\hat{\sigma} a\sigma_2 \in \bar{\Phi} & &\text{alphabet of } \hat{\sigma} \text{ has no overlap with } \sigma_1 a^2 \sigma_2
\end{aligned}$$

## 2  Notation

Let $|\sigma|$ and $\|\sigma\|$ be, respectively, the length of the sequence $\sigma$ and the number of distinct symbols in $\sigma$. We say that a sequence $\sigma = (\sigma_j)_{1 \le j \le |\sigma|}$ is a *subsequence* of $\Sigma = (\Sigma_j)_j$ if there exist $|\sigma|$ indices $j_1 < j_2 < \cdots < j_{|\sigma|}$ such that $\sigma_i = \Sigma_{j_i}$. Two sequences are *isomorphic* if they are identical up to renaming of symbols. We write $\sigma \prec \Sigma$ and $\sigma \prec \Sigma$ to mean, respectively, that $\sigma$ is a subsequent of $\Sigma$ and that $\sigma$ is *isomorphic* to a subsequence of $\Sigma$. A sequence $\Sigma$ (or class of sequences) is $\sigma$-*free* if $\sigma \not\prec \Sigma$. A sequence $\sigma = (\sigma_j)_j$ is $c$-*regular* if $\sigma_i = \sigma_j$ implies $|j - i| \ge c$. For instance, a 2-regular sequence has no immediate repetitions.

**Definition 2.1**  $\mathrm{Ex}(\sigma, n) = \max \{ |\Sigma| \,:\, \sigma \not\prec \Sigma \ \text{ and } \ \|\Sigma\| = n \ \text{ and } \ \Sigma \text{ is } \|\sigma\|\text{-regular} \}$

The condition that $\Sigma$ be $\|\sigma\|$-regular simply rules out uninteresting sequences. For instance, the infinite sequence $abababab\cdots$ is $(abc)$-free, but in the least interesting way.

A *symbol* refers to a member of the alphabet of a sequence and is distinguished from an occurrence of a symbol. For instance $abbccbbc$ contains 3 symbols, with 1, 4, and 3 occurrences of $a, b$, and $c$, respectively. In the text and figures we typically use the roman alphabet $a, b, c, \ldots$ but in the proofs it is more convenient to use the natural numbers $1, 2, 3, \ldots$.

If $u$ and $v$ are vertices in a rooted tree, $u \triangleleft v$ means that $u$ is a *strict* descendant of $v$, and $u \trianglelefteq v$ means $u \triangleleft v$ or $u = v$. A *generalized path compression* [12] is an operation that, given a sequence $(u_i, u_{i-1}, \ldots, u_1, u_0)$, where $u_i \triangleleft u_{i-1} \cdots \triangleleft u_1 \triangleleft u_0$, makes $u_i, \ldots, u_1$ the children of $u_0$ but otherwise does not affect the structure of the given tree. (This definition differs from a standard path compression, where $u_0$ is the parent of $u_1$, which is the parent of $u_2$, an so on.) We frequently call this operation a path compression or simply a compression. We say the compression *originates* at $u_i$ and *terminates* at $u_0$. The vertices $u_i, \ldots, u_1$ *participate* in the compression and the length of the compression is the number of participating vertices.

## 3  Path Compression Systems

Our construction of path compression systems follows the same lines as Hart and Sharir [12, 2] and Tarjan [25]. Given parameters $i, j$ we construct a complete binary tree $T(i, j)$ where nodes on each level of the tree are assigned an integer label. Based on this labeling we construct a sequence of $j \cdot |T(i, j)|$ path compressions, each with length $i$. The path compressions are then transcribed as a sequence $\Xi_{i,j}$ which avoids $abab$ as well as an infinite number of other forbidden subsequences. For any trees $T, T'$, $|T|$ is the number of leaves in $T$ and $T \circ T'$ (the composition of $T$ and $T'$) is derived by replacing each leaf of $T'$ with a copy of $T$.

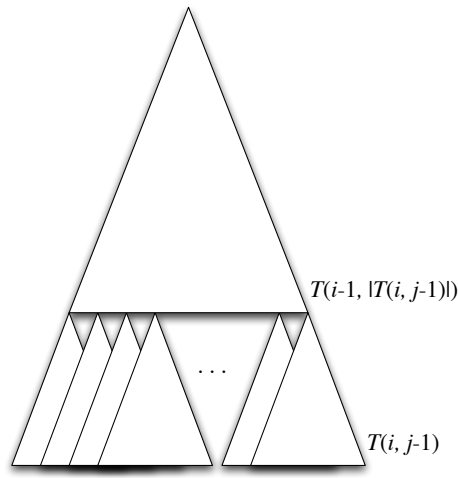

Figure 2: Composition of trees in the construction of $T(i, j)$.

Clearly $|T \circ T'| = |T| \cdot |T'|$; see Figure 2. Let $t_j$ be a full binary tree with $2^j$ leaves. The tree $T(i,j)$ is defined recursively, as follows:

$$\begin{aligned} T(1,j) &= t_j \\ T(i,0) &= t_c \qquad \{c \geq 1 \text{ is an arbitrary constant}\} \\ T(i,j) &= T(i,j-1) \circ T(i-1, |T(i,j-1)|) \end{aligned}$$

In other words, $T(i,j)$ is the composition of $t_c$ and $j$ trees of the form $T(i-1,\cdot)$, each of which is the composition of $t_c$ and several trees of the form $T(i-2,\cdots)$ and so forth. The leaves of a tree $T(i,\cdot)$ are called *i-nodes* and are ordered from left to right. The internal nodes of any $t_j$ are 0-nodes. If $u$ is a leaf of $T(i,j)$, $\nu_k(u)$ refers to the $k$th $(i-1)$-node ancestor of $u$ in $T(i,j)$, where $1 \leq k \leq j$. If $u$ is the $\ell$th leaf of $\nu_k(u)$ in $T(i,j)$ then $\mu_k(u) = \ell$. See Figure 3 for an illustration. We define a sequence of path compressions as follows.

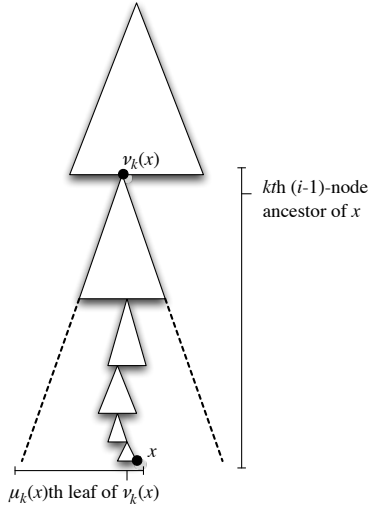

Figure 3: Illustration of $\nu_k(x)$ and $\mu_k(x)$.

Each leaf of $T(i,j)$ is the origin of $j$ path compressions and the path compressions are performed in postorder by their point of origin. Let $C(x,k)$ be the $k$th path compression originating from a leaf $x$ in $T(i,j)$. If $i=1$ let $C(x,k) = (x, \nu_k(x))$. For $i > 1$ let $C(x,k) = x \cdot C(\nu_k(x), \mu_k(x))$. We transcribe these path compressions into a sequence $\Xi_{i,j}$ as follows. Each path compression and each vertex is assigned a distinct symbol and we write $p \stackrel{c}{<} q$ if compression $p$ precedes compression $q$. Let $\xi(x)$ be the path compressions that $x$ participates in, listed in decreasing order by $\stackrel{c}{<}$. Let $\Xi_{i,j} = \xi(u_1) \cdot u_2 \cdot \xi(u_2) \cdot u_3 \cdot \xi(u_3) \cdots u_{2|T(i,j)|-1} \cdot \xi(u_{2|T(i,j)|-1})$, where $u_k$ is the $k$th vertex of $T(i,j)$ in postorder. The symbol $u_k$ appearing in $\Xi_{i,j}$ is distinct from the symbols of all path compressions. Since its only function is to enforce a regularity condition we will generally ignore these symbols. If $i$ and $j$ are understood or unimportant they will be omitted. For instance, $T$ refers to $T_{i,j}$ and the statements $\sigma \prec \Xi$ and $\sigma \not\prec \Xi$ should be interpreted as $\sigma \prec \Xi_{i,j}$ for some $i,j$ and $\sigma \not\prec \Xi_{i,j}$ for all $i,j$.

Lemma 3.1 just connects the length of $\Xi$ to the usual one- and two-argument versions of the inverse-Ackermann function. Its full proof is standard and tedious.

**Lemma 3.1** *Let $n = \|\Xi_{i,j}\|$ and $l = |T(i,j)|$. Then $|\Xi_{i,j}| = \Omega(n\alpha(n,l))$. When $j = O(1)$, $|\Xi_{i,j}| = \Omega(n\alpha(n))$*

Lemma 3.2 just lets us ignore regularity issues. For instance, we might say that $\sigma \not\prec \Xi$ without bothering to explicitly mention that $\Xi$ is $\|\sigma\|$-regular. Recall that $c$ is the constant from the definition of $T(i,0)$.

**Lemma 3.2** *For $i > 1$ or $j \geq c$, $\Xi_{i,j}$ is c-regular.*

**Proof:** Let $u, v$ be two vertices included in some path compression $a$, where $v$ is ancestral to $u$. One can see from the recursive construction of $T(i, j)$ and its associated path compressions that $u$ and $v$ are not 0-nodes and that they are at distance at least $c$. Therefore, there must be at least $c$ symbols between consecutive occurrences of $a$ in $\Xi$. □

Lemma 3.3 is invoked repeatedly in order to simplify proofs and obtain contradictions.

**Lemma 3.3** *For any two path compressions $p \overset{c}{<} q$, $qpqp \overset{\bar{c}}{\not\preceq} \Xi$.*

**Proof:** Let $u, v, w, x$ be the vertices associated with the respective occurrences of $q$ and $p$ in the purported subsequence $qpqp$ appearing in $\Xi$. (I.e., $\xi(u)$ and $\xi(w)$ contain $q$, $\xi(v)$ and $\xi(x)$ contain $p$.) It follows from the inequality $p \overset{c}{<} q$ that $u \trianglelefteq v \triangleleft w \trianglelefteq x$; see Figure 4. One effect of the $p$th path compression is to make $v$ and $x$ siblings and, as a direct consequence, to destroy the ancestor-descendant relationship between $u$ and $w$. Therefore no subsequent path compression can include both $u$ and $w$. Contradiction. □

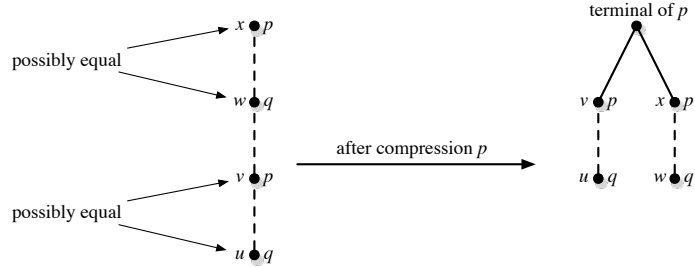

Figure 4: After compression $p$ no compression can include both $u$ and $w$.

Notice that Lemma 3.3 did not rely on any of the structure of $\Xi$; the proof would go through if $\Xi$ were the transcription of any system of path compressions. One corollary of Lemma 3.3 is that $\Xi$ is $(ababa)$-free, which, by Lemma 3.1, implies that $\mathrm{Ex}(ababa, n) = \Omega(n\alpha(n))$. This is one half of Hart and Sharir's proof [12] that $\mathrm{Ex}(ababa, n) = \Theta(n\alpha(n))$.

**Lemma 3.4** *Let $u, v, w$ be vertices with $u \triangleleft v \trianglelefteq w$ and suppose that $p \in \xi(v)$ and $q \in \xi(u), \xi(w)$. If the compression $p$ originates at a descendant of $u$ then $p \in \xi(u)$.*

**Proof:** First observe that for any compression $(u_i, u_{i-1}, \ldots, u_0)$ in our system, the intermediate vertices are uniquely determined by $u_i$ and $u_0$. In particular, $u_{i'}$ is the *first* $i'$ node on the path from $u_0$ to $u_i$. (In general any two vertices uniquely determine the whole compression.) Suppose that $u$ is an $i'$-node and $w$ an $i''$-node, where $i' > i''$; see Figure 5. It follows that $u$ is the first $i'$-node on the path from $w$ to the origin of compression $q$. Since $v \trianglelefteq w$ and compression $p$ originates below $u$, it also follows that $u$ is the first $i'$-node on the path from $v$ to the origin of $p$. Thus $p \in \xi(u)$. $\square$

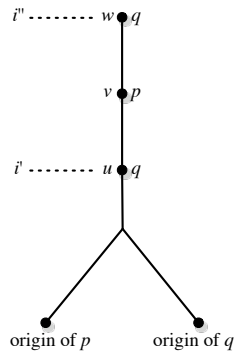

Figure 5: The situation that causes $p$ to make an *"implied"* appearance in $\xi(u)$.

Lemma 3.4 serves a simple purpose in our construction of forbidden subsequences. It says that aside from the explicit presence of $q \in \xi(u), \xi(w)$ and $p \in \xi(v)$, there must be an *implied* appearance of $p \in \xi(u)$. We deliberately design forbidden subsequences that are ($pqpqp$)-free but nonetheless cause implied symbols to appear in inconvenient places, leading to implied occurrences of $pqpqp$.

## 4 Encoding Forbidden Subsequences

Our most compact representation of forbidden subsequences is as strings over the alphabet $\{\heartsuit, \diamondsuit, \clubsuit, \spadesuit, \star\}$, where some groups of letters may be parenthesized. Each letter (or parenthesized set of letters) represents one symbol in the associated forbidden subsequence. The symbols corresponding to $\heartsuit, \diamondsuit, \clubsuit$, and $\spadesuit$ are called the *binder*, the *guard*, the *trap*, and the *trapped*. The roles of these symbols within the forbidden subsequence will be much easier to explain after analyzing a couple examples.

**Theorem 4.1** $\mathrm{Ex}(abcaccbc, n) = \Omega(n\alpha(n))$.

**Proof:** Suppose that $\sigma = abcaccbc$ were to occur in $\Xi$. By Lemma 3.3 we can eliminate all cases except $a \stackrel{c}{<} b \stackrel{c}{<} c$. Let $v_{a,i}$ be the vertex in $T$ corresponding to the $i$th occurrence of $a$ in $\sigma$. It follows from the construction of $\Xi$ that $v_{a,1}, v_{b,1}, v_{c,1} \trianglelefteq v_{a,2}$ and that $v_{c,2} \triangleleft v_{c,3} \trianglelefteq v_{b,2} \triangleleft v_{c,4}$. See Figure 6. We apply Lemma 3.4 to the symbols $b$ and $c$ occurring in $\xi(v_{c,2}), \xi(v_{b,2})$, and $\xi(v_{c,4})$ and conclude that $b$ must also appear in $\xi(v_{c,2})$. In other words, if $abcaccbc$ appears in $\Xi$ then $abcac\underline{b}cbc$ appears as well. Since, by Lemma 3.3, $\Xi$ contains no subsequences isomorphic to $ababa$, it must also be $\sigma$-free. Therefore, $\mathrm{Ex}(\sigma, n) \geq |\Xi| = \Omega(n\alpha(n))$. $\square$

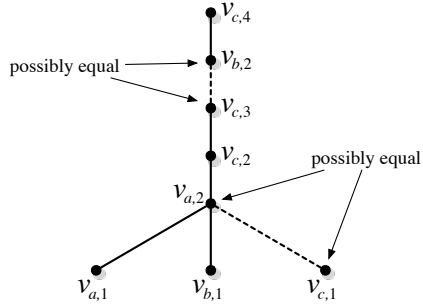

Figure 6: The vertex $v_{x,i} \in T$ is such that $\xi(v_{x,i})$ contains the symbol corresponding to the $i$th occurrence of $x$ in $\sigma$. Dashed lines connect vertices that may be the same.

Let us analyze the functions of $a, b$, and $c$ in the proof of Theorem 4.1. The symbol $a$ did not appear in the ultimate contradiction (the implied subsequence $bcbcbc$) but it did facilitate the contradiction by forcing $v_{b,1}$ and $v_{c,1}$ to be descendants of $v_{c,2}, v_{c,3}, v_{b,2}$, and $v_{c,4}$. In our terminology $a$ is the *binder* (symbolized by $\heartsuit$) because it binds previous symbols (i.e., vertices in $T$) under one common ancestor. The locations of $b$ and $c$ were chosen with the preconditions of Lemma 3.4 in mind. For the proof to go through we need $c$ to appear in $\xi(v_{c,2})$ and $\xi(v_{c,4})$ and $b$ to appear in $\xi(v_{b,2})$, and, crucially, that $v_{c,2} \triangleleft v_{b,2}$. This last condition is enforced by the immediate repetition of $c$ in $\sigma$. In our terminology $c$ acts as a *guard* (making sure $v_{b,2}$ is a strict ancestor of $v_{c,2}$) and both $b$ and $c$ are *trapped* by $c$, meaning that the symbols $b$ and $c$ appear at vertices that lie strictly above one occurrence of $c$ and strictly below another occurrence of $c$. Guards, traps, and trapped symbols are represented by $\diamondsuit, \clubsuit$, and $\spadesuit$, respectively. Thus, we can represent $\sigma$ as $\heartsuit\spadesuit(\diamondsuit\spadesuit\clubsuit)$: $a$ acts as a binder, $b$ as a trapped symbol, and $c$ as a guard, a trap, and a trapped symbol. It is not true that *every* string over $\{\heartsuit, \diamondsuit, \spadesuit, \clubsuit, \star\}$ can be realized as a new superlinear forbidden subsequence. However, with only a few syntactic restrictions on the encoding we can show that each valid encoding corresponds to at most a constant number of forbidden subsequences, each of which is independent of the others. Before giving these restrictions we look at one more specific example.

**Theorem 4.2** $\text{Ex}(abcbdadbcd, n) = \Omega(n\alpha(n))$.

**Proof:** As before, suppose that $\sigma = abcbdadbcd \prec \Xi$. It follows from Lemma 3.3 that $a \stackrel{c}{<} b \stackrel{c}{<} c \stackrel{c}{<} d$ and from the construction of $\Xi$ that $v_{a,1}, v_{b,2}, v_{d,1} \trianglelefteq v_{a,2}$, that $v_{b,1}, v_{c,1} \trianglelefteq v_{b,2}$, and that $v_{d,2} \triangleleft v_{c,2} \triangleleft v_{d,3}$. Lemma 3.4 (applied to $c$ and $d$) implies that $c$ appears in $\xi(v_{d,2})$; another application of Lemma 3.4 (now with $c$ and $b$) implies that $c$ also appears in $\xi(v_{b,2})$; see Figure 7. In other words, if $\sigma$ appears in $\Xi$ then $\sigma' = abcbdad\underline{c}bcd$ appears in $\Xi$ as well. This leads to a contradiction since $bcbcbc \prec \sigma'$, which, by Lemma 3.3, can never appear in $\Xi$. □

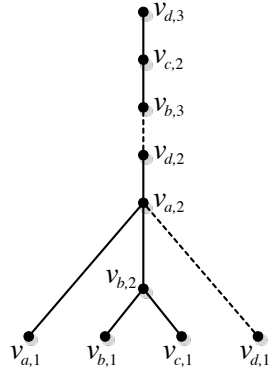

Figure 7: Dashed lines connect vertices that may be the same.

In $\sigma$ the appearance of $c$ in $\xi(v_{c,2})$ is trapped by $d$ and the implicit appearance of $c$ in $\xi(v_{d,2})$ is trapped by $b$. The symbol $a$ acts as a binder to insure that $v_{b,2} \triangleleft v_{d,2}$. The third occurrence of $b$ in $\sigma$ acts as a guard to ensure that $v_{d,2} \triangleleft v_{c,2}$ and the second occurrence of $d$ ensures that $v_{b,2} \triangleleft v_{d,2}$. We could encode $\sigma$ succinctly as $\heartsuit(\diamondsuit\clubsuit)\spadesuit(\diamondsuit\clubsuit)$ but it turns out that when there are two traps (by $d$ and $b$ in this case) they each act as guards for the other. There is no ambiguity in coding $\sigma$ as $\heartsuit\clubsuit\spadesuit\clubsuit$.

Notice that $\sigma$ is very similar but shorter than the forbidden subsequence considered by Klazar [17] : $\sigma' = abcbadadbcd$. The relevant difference is that $G(\sigma')$ is strongly connected, which implies the superlinearity of $\text{Ex}(\sigma', n)$ [17], whereas $G(\sigma)$ is not. Thus, the validity of Theorem 4.2 follows from different principles.

Definition 4.3 uses the following regular expression notation: $X^*$ represents zero or more repetitions of $X$ and $[X, Y, Z]$ represents exactly one of $X, Y,$ and $Z$.

**Definition 4.3** *A string in* $\{\heartsuit, \diamondsuit, \spadesuit, \clubsuit, \star, (,)\}^*$ *is a* legal compact encoding *if it is*

    (1)   $\heartsuit\clubsuit\spadesuit\clubsuit$                                  (8–10)   *in* $\star\spadesuit\star^*\heartsuit[(\diamondsuit\spadesuit\clubsuit),(\diamondsuit\spadesuit)\clubsuit,\diamondsuit\spadesuit\clubsuit]$

    (2–4)   $\heartsuit\spadesuit[(\diamondsuit\spadesuit\clubsuit),(\diamondsuit\spadesuit)\clubsuit,\diamondsuit\spadesuit\clubsuit]$       (11–12)   *in* $\star\diamondsuit\spadesuit\star^*\spadesuit\star^*\heartsuit\clubsuit$ *or* $\diamondsuit\star\spadesuit\star^*\spadesuit\star^*\heartsuit\clubsuit$

    (5)   $\heartsuit\diamondsuit\spadesuit\spadesuit\clubsuit$                              (13–14)   *in* $\star\diamondsuit\spadesuit\star^*\heartsuit\spadesuit\clubsuit$ *or* $\diamondsuit\star\spadesuit\star^*\heartsuit\spadesuit\clubsuit$

    (6)   $\diamondsuit\heartsuit\spadesuit\spadesuit\clubsuit$                              (15–16)   *in* $\star\spadesuit\star^*[(\diamondsuit\spadesuit),\diamondsuit\star^*\spadesuit]\star^*\heartsuit\clubsuit$

    (7)   *in* $\star\clubsuit\spadesuit\star^*\heartsuit\clubsuit$                 (17)   *in* $\star\spadesuit\star^*\diamondsuit\star^*\heartsuit\spadesuit\clubsuit$

The legal strings from Definition 4.3 could be generated from an alternative set of rules which are equally unintuitive but may be helpful to keep in mind while reading the proofs.

**Definition 4.4** *A string is a legal compact encoding if it is of the form given in Definition 4.3(1,7) or if it contains exactly one $\heartsuit$, $\diamondsuit$, and $\clubsuit$, two $\spadesuit$s, possibly an unbounded number of $\star$s, at most one parenthesized expression which is either $(\diamondsuit\spadesuit)$ or $(\diamondsuit\spadesuit\clubsuit)$, and is subject to the following restrictions:*

    (i)   *The final symbol must be* $\clubsuit$                (iv)   *A $\star$ or the $\heartsuit$ must precede both $\spadesuit$s*

    (ii)   *The $\diamondsuit$ must precede at least one $\spadesuit$*       (v)   *A $(\diamondsuit\spadesuit)$ cannot precede the other $\spadesuit$*

    (iii)   *All $\star$s must precede the $\heartsuit$*                 (vi)   *The first two symbols that are either $\star$ or $\heartsuit$ must have a $\spadesuit$ between them*

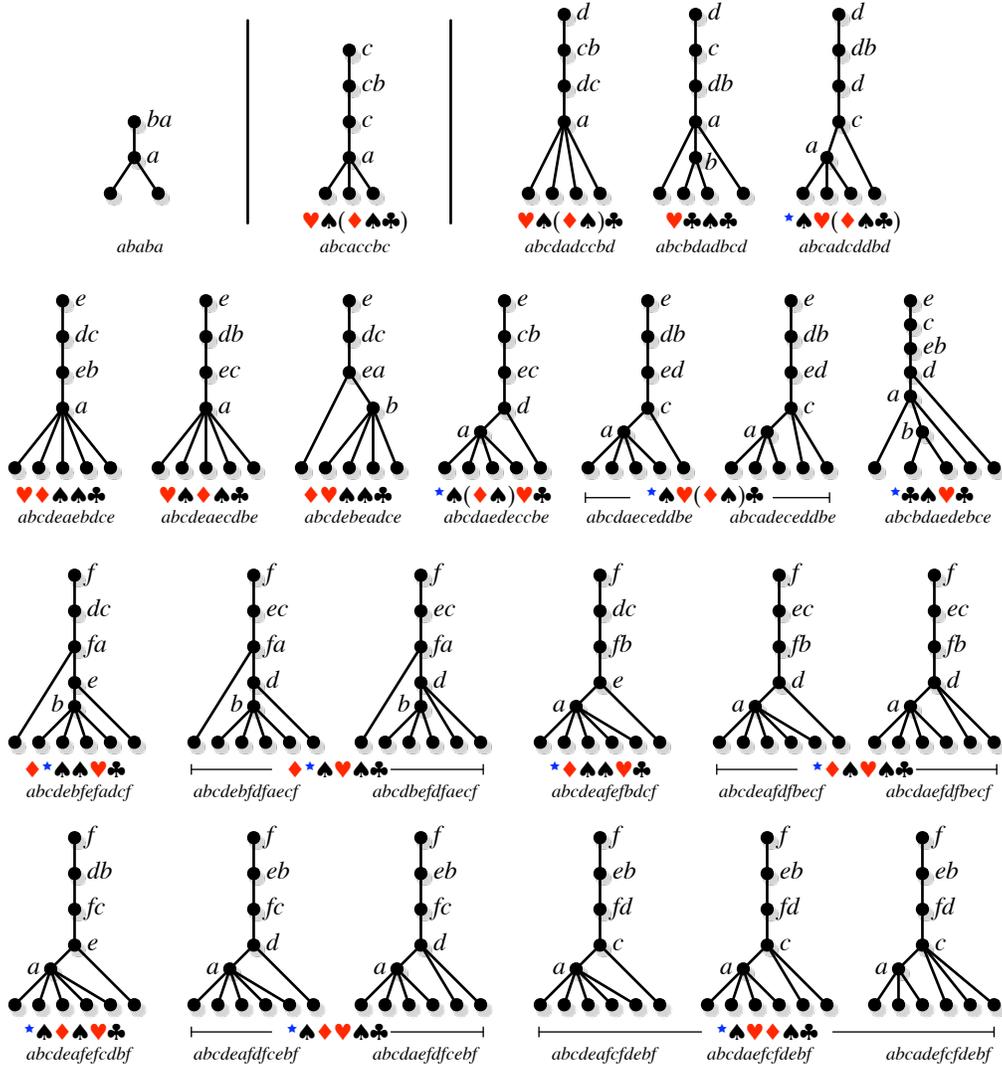

Figure 8: The 17 prototypes. Each prototype has a compact encoding (using the minimum number of ⋆s) and each encoding can be realized by at most a constant number of labeled trees, each of which corresponds to a forbidden subsequence. The vertex labels are indicated, except for the leaves. In each tree the leaves are labeled in left-to-right order: $a, b, c, d, \ldots$.

Most *illegal* strings still lead unambiguously to forbidden subsequences. We designate them illegal either because we cannot prove that they are superlinear, or because they are trivially superlinear, e.g., if they contain *ababa* or another subsequence known to be superlinear.

We describe below a two step process for converting a legal compact encoding from Definition 4.3 into a forbidden subsequence. The first step converts a compact encoding into a labeled tree representation. There may be more than one possible tree representation per compact encoding, though never more than four. The second step is to map a labeled tree into a forbidden subsequence; this mapping is always unique. See Figure 8 for a depiction of the 17 prototypes. Each is represented as a legal encoding, a set of 1 or more trees, and for each tree a corresponding forbidden subsequence. More complex forbidden subsequences can be derived by inserting ⋆s into the compact encoding.

Let $\lambda$ be a legal compact encoding. An *element* is a symbol in $\{\heartsuit, \diamondsuit, \spadesuit, \clubsuit, \star\}$ or a parenthesized sequence of symbols from that set. Let $|\lambda|$ be the number of elements in $\lambda$ and $\lambda(j)$ be the $j$th element. We generate a tree $\tau_\lambda$ from the bottom up as follows. We create $|\lambda|$ leaves where the $j$th leaf $\ell_j$ is labeled $j$. (If $\lambda$ contains consecutive elements $\lambda(j)\lambda(j+1) = \clubsuit\spadesuit$, as in cases (1) or (7) from Definition 4.3, we create a new node $\ell'_j$ labeled $j$ that is the parent of $\ell_j$ and $\ell_{j+1}$. If this is the case, substitute $\ell'_j$ for all references to $\ell_j$ and $\ell_{j+1}$ below.) Suppose $\lambda(j_1), \ldots, \lambda(j_{l-1})$ are the '$\star$' elements and $\lambda(j_l)$ the '$\heartsuit$' element of $\lambda$. We create $l$ new internal nodes $y_1, \ldots, y_l$, where $y_i$ is labeled $j_i$. We make $y_1$ the parent of $\ell_{j_1}$ and $\ell_{j_2}$, and in general, $y_i$ is made the parent of $y_{i-1}$ and $\ell_{j_{i+1}}$. Finally, $y_l$ is made the parent of $\ell_{|\lambda|}$. Figure 9 shows $\tau_\lambda$ for $\lambda = \star\spadesuit \star^2 \diamondsuit \star^2 \spadesuit\heartsuit\clubsuit$.

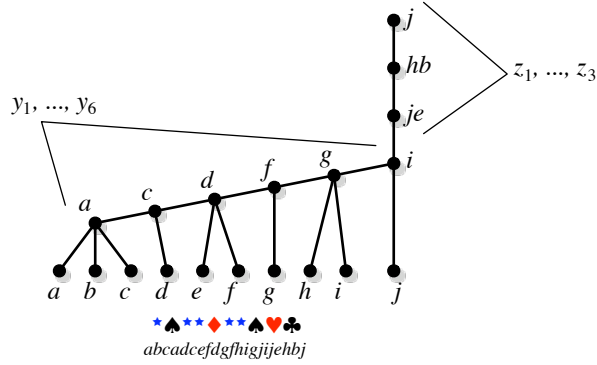

Figure 9: One example $\tau_\lambda$, for $\lambda = \star\spadesuit \star^2 \diamondsuit \star^2 \spadesuit\heartsuit\clubsuit$.

If a leaf $\ell_{j'}$ has siblings to the left and right that are the children of a common parent $y_i$ then $\ell_{j'}$ must also be a child of $y_i$. Some leaves may be able to choose between one of two parents. For instance, if $\lambda = \star\diamondsuit\spadesuit\heartsuit\spadesuit\clubsuit$ then $y_1$ is the parent of $\ell_1$ and $\ell_4$, which forces $\ell_2$ and $\ell_3$ to be children of $y_1$ as well. We put $y_2$ as the parent of $y_1$ and $\ell_6$, and have the freedom to make $\ell_5$ the child of either $y_1$ or $y_2$. See the last two trees on the third row of Figure 8. We create three nodes ancestral to $y_l$ named $z_1, z_2, z_3$. Let $j^\diamondsuit$ be the index of the guard element of $\lambda$, $j_1^\spadesuit < j_2^\spadesuit$ be the indices of the trapped elements, and $j_1^\clubsuit < j_2^\clubsuit$ the indices of the traps. (It is impossible for all these elements to be present simultaneously.) If $\lambda$ contains two traps (see cases (1) and (7) of Definition 4.3) we assign $z_1$ label $j_2^\clubsuit j_1^\clubsuit$, $z_2$ label $j_1^\spadesuit$ (there is no $j_2^\spadesuit$), and $z_3$ label $j_2^\clubsuit$. If $\lambda$ contains just one trap, namely $j_1^\clubsuit$, we assign $z_2$ label $j_2^\spadesuit j_1^\spadesuit$, $z_3$ label $j_1^\spadesuit$, and $z_1$ label $j_1^\clubsuit j^\diamondsuit$, unless $j_1^\clubsuit = j^\diamondsuit$, in which case $z_1$ is simply labeled $j_1^\clubsuit$. See Figure 8 for the labeled tree representations of the prototypes.

Once we settle on a particular labeled tree $\tau$, turning it into a sequence is relatively straightforward. Let $\sigma(\tau)$ be the concatenation of the labels of the nodes of $\tau$ in the (unique) postorder in which $\ell_1$ precedes $\ell_2$, which precedes $\ell_3$, and so on.

**Definition 4.5** $\Psi$ *is the set containing abab and all sequences that can be generated from a legal compact encoding.*

## 5 General Superlinear Lower Bounds

In this Section we prove our main result, that all forbidden subsequences that could be generated from a legal compact encoding are superlinear.

**Theorem 5.1** *For all $\sigma \in \Psi$, $\text{Ex}(\sigma, n) = \Omega(n\alpha(n))$*

**Proof:** Let $\lambda$ be the legal compact encoding that generated $\sigma$. Theorems 4.1 and 4.2 cover the case of $\lambda \in \{\heartsuit\spadesuit(\diamondsuit\spadesuit\clubsuit), \heartsuit\clubsuit\spadesuit\clubsuit\}$, i.e., parts (1) and (2) of Definition 4.3. It is straightforward, given the proof of Theorem 4.2 and what follows, to cover $\lambda \in \star\clubsuit\spadesuit \star^* \heartsuit\clubsuit$. For notational simplicity we will omit this case. Thus $\lambda$ contains exactly one $\clubsuit, \diamondsuit$, and $\heartsuit$, two $\spadesuit$s, an unbounded number of $\star$s, and at most one parenthesized element, which is either $(\diamondsuit\spadesuit)$ or $(\diamondsuit\spadesuit\clubsuit)$. (It may be helpful at this point to reread the properties of legal encodings from Definition 4.4.) Let $\lambda(j_1), \ldots, \lambda(j_{l-1})$ be the $\star$ elements, $\lambda(j_l)$ the $\heartsuit$, and $\lambda(j^\diamond), \lambda(j_1^\spadesuit), \lambda(j_2^\spadesuit)$, and $\lambda(j^\clubsuit)$ the elements of type $\diamond, \spadesuit$ and $\clubsuit$, respectively. Observe that if $\lambda$ contains no parenthesized expressions, $j^\clubsuit$ appears three times in $\sigma$ and all other symbols appear exactly twice. If there is a parenthesized expression, i.e., $j^\diamond = j_2^\spadesuit$ or $j^\diamond = j_2^\spadesuit = j^\clubsuit$, then that symbol will appear three and fours times in $\sigma$, respectively.

Suppose that $\sigma \prec \Xi$. Let $v_{j,i}$ be the vertex in $T$ such that $\xi(v_{j,i})$ contains the $i$th occurrence of $j$ in $\sigma$. In order to show that this leads to a contradiction we first need to show that the induced subtree of $T$ tree connecting $\{v_{j,i}\}$ mimics the structure of $\tau_\lambda$. Assume inductively that $v_{j_k,2}$ is ancestral to $v_{j_1,1}, v_{j_{1+1},1}, \ldots, v_{j_{k+1},1}$. Following the 2nd occurrence of $j_k$ in $\sigma$ we see the first occurrence of some symbols that must include $j_{k+2}$, followed by the second occurrence of $j_{k+1}$. Since $v_{j_{k+1},1} \lhd v_{j_{k+1},2}$ (trivial) and $v_{j_{k+1},1} \lhd v_{j_k,2}$ (ind. hyp.), it follows from the construction of $\Xi$ that $v_{j_k,2} \lhd v_{j_{k+1},2}$. This implies, by the inductive hypothesis, that $v_{j_1,1}, \ldots, v_{j_{k+2},1}$ are descendants of $v_{j_{k+1},2}$. Furthermore, $v_{j_1,1}, \ldots, v_{j^\clubsuit}$ are descendants of $v_{j_l,2}$. In other words, the symbol $j_l$ (corresponding to the $\heartsuit$ in $\lambda$) functions as intended: by binding $v_{j_1^\spadesuit,1}, v_{j_2^\spadesuit,1}$, and $v_{j^\clubsuit,1}$ under a common ancestor. The next step is to show that $j^\diamond$ functions as a guard.

The symbols following the second occurrence of $j^\clubsuit$ in $\sigma$ are $j^\diamond, j_2^\spadesuit, j_1^\spadesuit$, and $j^\clubsuit$ in that order, where $j^\diamond$ is omitted if $j^\diamond = j^\clubsuit$ (as in cases (2) and (8) of Definition 4.3.) First consider the case where $j^\diamond, j^\clubsuit$, and $j_2^\spadesuit$ are distinct. Regardless of the compact encoding $\lambda$, it always holds that $j^\diamond j^\clubsuit j^\diamond j^\clubsuit$ and $j^\diamond j_2^\spadesuit j^\diamond j_2^\spadesuit$ appear in $\sigma$; see Definition 4.4(i,ii). Thus, it follows from Lemma 3.3 that $j^\diamond \stackrel{c}{<} j^\clubsuit$ and $j^\diamond \stackrel{c}{<} j_2^\spadesuit$. Since $\xi(v_{j^\clubsuit,2})$ appears in $\Xi$ in *decreasing* order (by $\stackrel{c}{<}$), it must be that $v_{j^\clubsuit,2} \lhd v_{j_2^\spadesuit,2}$. This also implies that $v_{j^\clubsuit,2} \lhd v_{j_1^\spadesuit,2} \unlhd v_{j^\clubsuit,3}$. The preconditions of Lemma 3.4 are satisfied (with respect to both of the pairs $j^\clubsuit, j_1^\spadesuit$ and $j^\clubsuit, j_2^\spadesuit$), implying that $j_1^\spadesuit$ and $j_2^\spadesuit$ appear in $\xi(v_{j^\clubsuit,2})$. Therefore, if $\sigma \prec \Xi$ then $j_1^\spadesuit j_2^\spadesuit \underline{\mathbf{j_1^\spadesuit}} j_2^\spadesuit j_1^\spadesuit \prec \Xi$ as well, contradicting Lemma 3.3.

If $j^\diamond = j^\clubsuit = j_2^\spadesuit$ or $j^\diamond = j_2^\spadesuit$ (the cases where $\lambda$ contains $(\diamond\spadesuit\clubsuit)$ and $(\diamond\spadesuit)$ respectively) the above proof goes though with somewhat simpler arguments. if $j^\diamond = j_2^\spadesuit$ then $v_{j_2^\spadesuit,2} \lhd v_{j_2^\spadesuit,3}$ (this is trivial) and by Lemma 3.4, $j_1^\spadesuit$ appears in $\xi(v_{j^\clubsuit,2})$. Thus, if $\sigma \prec \Xi$ then $j_1^\spadesuit j_2^\spadesuit \underline{\mathbf{j_1^\spadesuit}} j_2^\spadesuit j_1^\spadesuit \prec \Xi$ as well. The cases where $j^\diamond = j^\clubsuit = j_2^\spadesuit$ are treated similarly. □

A natural question is whether there is any redundancy in $\Psi$. That is, if $\sigma, \hat\sigma \in \Psi$ and $\hat\sigma$ is a strictly shorter subsequence of $\sigma$, then the superlinearity of $\mathrm{Ex}(\hat\sigma, n)$ (Theorem 5.1) immediately implies the superlinearity of $\mathrm{Ex}(\sigma, n)$. If the superlinearity of $\Psi$ could be deduced from a strict subset (perhaps even a finite subset), this would undermine the claim that the infinitude of $\Psi$ is evidence for the infinitude of $\Phi$. Theorem 5.2 shows that $\Psi$ is minimal in the sense that every strict subsequence of an element in $\Psi$ *does* appear as a subsequence of $\Xi$.

**Theorem 5.2** *For any $\sigma \in \Psi$, if $\hat\sigma \prec \sigma$ and $\hat\sigma \neq \sigma$ then $\hat\sigma \prec \Xi$.*

**Proof:** There are seventeen prototypes and, unfortunately, we have no elegant way to capture all of them with a single argument. Note, however, that prototypes (1–6) are fixed length and can be checked by hand. Prototype (7) is an oddball; however, it is simple to handle given what follows. The bulk of the proof covers cases (10–14, 16–17), which are the remaining ones that contain no parenthesized expressions. We sketch how to handle prototypes (8–9, 15) at the end.

Let $\lambda$ be the compact encoding for $\sigma$. Let $j_1, \ldots, j_l$ be the indices in $\lambda$ of the $\star$s and the $\heartsuit$; let $j^\diamond, j_1^\spadesuit, j_2^\spadesuit$, and $j^\clubsuit$ be the indices of their respective types and assume for the moment that all these indices are distinct, i.e., assume there is no $(\diamond\spadesuit)$ or $(\diamond\spadesuit\clubsuit)$ in $\lambda$. Let $\hat\sigma$ be a strict subsequence of $\sigma$. If $\hat\sigma$ is missing a symbol corresponding to a $\star, \diamond, \heartsuit$, or $\spadesuit$ we can assume without loss of generality that the other occurrence of that symbol is missing as well.

First consider the case where $\hat\sigma$ is missing both occurrences of $j_h$, for some $1 \leq h \leq l$. We show that the $\heartsuit$ element no longer functions as a binder. As before, let $v_{j,i} \in T$ be the vertex corresponding to the $i$th occurrence of $j$ in the purported appearance of $\hat\sigma \prec \Xi$. Unless explicitly contradicted, assume that (i) $v_{j,1}$ is a leaf of $T$, (ii) all $v_{j,i}$ are distinct, and (iii) $v_{j,i} \lhd v_{j',i'}$ only if this relation must hold, given $\hat\sigma$. (For instance, if $\hat\sigma = abab$, (iii) would require that $v_{a,2} \lhd v_{b,2}$.) Let $\hat\tau$ be a tree on the vertices $\{v_{j,i}\}$ modeling the necessarily the ancestor/descendant relationships of (iii). In order to prove the theorem we only need to assign the $v_{j,i}$ to levels in $T$ that is consistent with $\hat\tau$ and the construction of path compressions from Section 3.

Let $\sigma = \sigma_1 \, j_h \, \sigma_2 \, j_{h-1} \, \sigma_3 \, j_{h+1} \, \sigma_4 \, j_h \, \sigma_5 \, j^\diamond \, \sigma_6$, where the $j^\diamond$ refers to its second occurrence in $\sigma$. If $h = l$ then $j_{h+1}$ refers to $j^\clubsuit$ and if $h = 1$, $j_{h-1}$ does not exist. Thus, $\hat{\sigma} = \sigma_1 \, \sigma_2 \, j_{h-1} \, \sigma_3 \, j_{h+1} \, \sigma_4 \, \sigma_5 \, j^\diamond \, \sigma_6$. A case-by-case check of Definition 4.3 (or the more readable Definition 4.4) shows that the alphabets of the sequences $\sigma_1 \, \sigma_2 \, j_{h-1}$, $\sigma_3$, and $j_{h+1} \, \sigma_4 \, \sigma_5$ are mutually disjoint. Furthermore, the first and second occurrences of $j^\clubsuit$ appear in $j_{h+1} \, \sigma_4 \, \sigma_5$ and $j_1^\spadesuit$ does not. Thus, in $\hat{\tau}$, the least common ancestor of $v_{j_1^\spadesuit, 1}$ and $v_{j^\clubsuit, 2}$ appears at or above $v_{j^\diamond, 2}$. Figure 10 illustrates the difference between $\tau$ and $\hat{\tau}$ on a specific example. Suppose that $T = T(k, \ell)$, so all leaves (of $T$ and $\hat{\tau}$) are $k$-nodes. Let $v_{j^\clubsuit, 2}$ be a $(k-1)$-node. For all compressions $j$ appearing in $\sigma_1 \, \sigma_2 \, j_{h-1} \, \sigma_3$, let the $(k-1)$-node of compression $j$ lie at the same level as $v_{j^\clubsuit, 2}$ and lie between $v_{j,1}$ and the parent of $v_{j,1}$ in $\hat{\tau}$. See Figure 10 for a schematic of how nodes in $\hat{\tau}$ are assigned to levels in $T$. For each such $j$, $v_{j,2}$ is a $(k-2)$-node and $v_{j^\clubsuit, 3}$ is a $(k-2)$-node. If $j^\diamond$ and/or $j_2^\spadesuit$ appear in $\sigma_4 \, \sigma_5$ then (by Lemma 3.4), $j^\diamond, j_2^\spadesuit$ appear in $\xi(v_{j^\clubsuit, 2})$ as well. Figure 10 gives an example of this situation: $h = j_2^\spadesuit$ appears in $\xi(v_{j,2}) = \xi(v_{j^\clubsuit, 2})$. In this case $v_{j^\clubsuit, 2}$ is the $(k-1)$-node of compressions $j^\diamond, j_2^\spadesuit$ and $v_{j^\diamond, 2}$ and $v_{j_2^\spadesuit, 2}$ are the $(k-2)$-nodes in compressions $j^\diamond$ and $j_2^\spadesuit$. For other compressions $j$ appearing in $\sigma_4 \, \sigma_5$ ($\star$s and $\heartsuit$), the nodes $v_{j,2}$ are $(k-1)$ nodes at distinct levels in $T$. (In Figure 10 $g$ and $i$ are in this category.) It is clear, i.e., as clear as anything else in this proof, that the existence of these compressions could not cause any contradictions since they are effectively sequestered from the others; we ignore them in the arguments below.

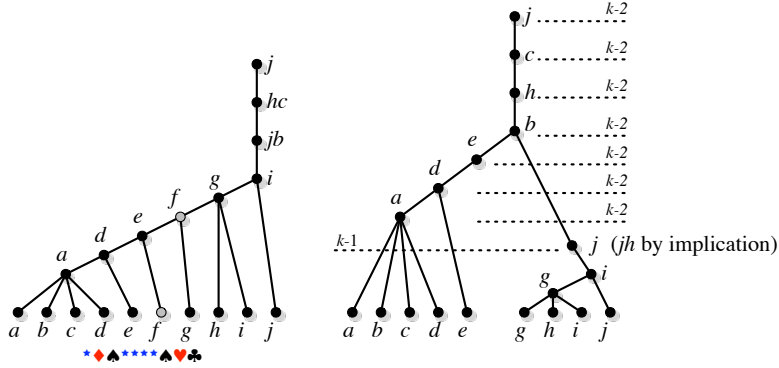

Figure 10: Left: the model tree corresponding to $abcdaedfegfhigjijbhcj$. Right: the model tree corresponding to $abcdaedeghigjijbjcj$, where '$f$' is missing.

Recapping the above, all compressions in $\sigma_1 \, \sigma_2 \, j_{h-1} \, \sigma_3, j^\diamond, j_2^\spadesuit, j^\clubsuit$ have their $(k-1)$-nodes at the same level and all these nodes are distinct, except possibly for $j^\diamond$ and $j_2^\spadesuit$, whose $(k-1)$-nodes may be equal to that of $j^\clubsuit$. For any $(k-1)$-node $x$ there is exactly one compression containing this node and each $(k-2)$-node ancestor of $x$ that lies below the first $(k-1)$-node ancestor of $x$. Thus, so long as the $(k-1)$-nodes of compressions are distinct, *any* permutation of them in the sequence $j^\diamond \sigma_6$ is consistent with our construction of path compressions. We only need to worry about the compressions that may share a common $(k-1)$-node. Since $j^\diamond < j_2^\spadesuit < j^\clubsuit$ and these compressions appear in sorted order in $j^\diamond \sigma_6$, i.e., $j^\clubsuit$ terminates above $j_2^\spadesuit$, which terminates above $j^\diamond$, these compressions are also consistent with our construction.

Consider now the case when $\hat{\sigma}$ is missing both occurrences of $j^\diamond$. Without the guard there is no contradiction in letting $v_{j_2^\spadesuit, 2} = v_{j_1^\spadesuit, 2} = v_{j^\clubsuit, 2}$. For every compression $j$ we let $v_{j,2}$ be the $(k-1)$-node of $j$ and let $v_{j^\clubsuit, 3}$ be the $(k-2)$-node of $j^\clubsuit$. All the $(k-1)$-nodes are distinct, with the exception of $v_{j_2^\spadesuit, 2} = v_{j_1^\spadesuit, 2} = v_{j^\clubsuit, 2}$. The same observation made above shows that all compressions with distinct $(k-1)$-nodes are consistent with our construction, independent of the locations of their corresponding appearances in $\hat{\sigma}$. Since $j_1^\spadesuit < j_2^\spadesuit < j^\clubsuit$ It is also consistent with our construction of $\Xi$ that $\xi(v_{j^\clubsuit, 2})$ list $j_1^\spadesuit, j_2^\spadesuit$, and $j^\clubsuit$ in decreasing order.

The cases where $\hat{\sigma}$ is missing $j_1^\spadesuit$ or $j_2^\spadesuit$ are similar to the case of $j^\diamond$ above. If $j_2^\spadesuit$ is missing there is no inconsistency in letting $v_{j^\clubsuit, 2}$ be the $(k-1)$-node in compression $j_1^\spadesuit$ and $v_{j_1^\spadesuit, 2}$ be the $(k-2)$-node in $j_1^\spadesuit$. As before, $v_{j^\clubsuit, 3}$ is the $(k-2)$-node of $j^\clubsuit$ and all other nodes and their labels are the same as if $j^\diamond$ were missing. Since $j_1^\spadesuit < j^\clubsuit$ there is no inconsistency in letting $j^\diamond, j_1^\spadesuit$, and $j^\clubsuit$ appear in $\xi(v_{j^\clubsuit, 2})$ in decreasing order.

Suppose that one of the three occurrences of $j^{\clubsuit}$ is missing. There are now exactly two occurrences of each symbol in $\hat{\sigma}$. For each $j$ let $v_{j,1}$ be the $k$-node of compression $j$ and $v_{j,2}$ be its $(k-1)$-node. We put all the $k$-nodes at the same level in $T$ whereas all of the $(k-1)$-nodes are at different levels. By our construction of path compressions there exists exactly one compression that includes $v_{j,1}$ and a particular $(k-1)$-node ancestor. Thus, regardless of arrangement of symbols in $\hat{\sigma}$ it is always possible to assign, in a consistent fashion, symbols to path compressions. (Note that the proof of this case goes through if all symbols appear in $\hat{\sigma}$ in at most two runs, e.g., *abbccccbbaaccc* has this property.)

Extending the proof above to cases where $\lambda$ may contain elements $(\diamondsuit\spadesuit)$ or $(\diamondsuit\spadesuit\clubsuit)$ is not difficult. In the first case the symbol corresponding to $(\diamondsuit\spadesuit)$ appears three times. For instance, if $\lambda = \heartsuit\spadesuit(\diamondsuit\spadesuit)\clubsuit$ (prototype (3)) the forbidden subsequence $\sigma = abcdadccbd$ contains three occurrences of $c$, corresponding to the $(\diamondsuit\spadesuit)$; see Figure 8. The first and third occurrences serve the role of a trapped element and the first and second serve the role of a guard. If $\hat{\sigma}$ is missing the second or third occurrence we use the same analysis as above, as if $j^{\diamondsuit}$ were missing. If $\hat{\sigma}$ is missing the first occurrence then the remaining two are consecutive in $\hat{\sigma}$; clearly their presence could not assist in obtaining a contradiction. The other cases, where $\hat{\sigma}$ is missing a $\star, \heartsuit, \spadesuit$, or $\clubsuit$, are handled in the same way as before. Turning to the case where $\lambda$ contains $(\diamondsuit\spadesuit\clubsuit)$, the corresponding symbol appears four times. For instance, the forbidden subsequence corresponding to $\heartsuit\spadesuit(\diamondsuit\spadesuit\clubsuit)$ is $\sigma = abcaccbc$, where the first, second, and fourth occurrences serve the role as a trap and the first and third occurrences serve as both a guard and a trapped element. If $\hat{\sigma}$ is missing the second or third occurrence we use the earlier analysis as if $j^{\diamondsuit}$ were missing. If $\hat{\sigma}$ is missing the first or fourth occurrence we use the analysis as if $j^{\clubsuit}$ were missing.

The arguments above cover all prototypes except (1) and (7), which are easy exercises.
□

Theorems 5.1 and 5.2 together show that there are an infinite number of superlinear forbidden subsequences, each of which is not a subsequence of any other. With the exception of *ababa* we cannot say that any particular member of $\Psi$ is in $\Phi$. However, we can show that $|\Phi| \geq 5$ non-constructively. Recall that the previous bound of $|\Phi| \geq 2$ [17] followed from the superlinearity of $\sigma' = ababa$ and $\sigma'' = abcbadadbcd$. Clearly $\sigma'$ is a palindrome (a sequence isomorphic to its reversal) and $\sigma''$ is not. If $\sigma''$ were in $\Phi$ its reversal would be there as well. However, since $\sigma''$ contains a palindrome as a subsequence, namely *abadadbd*, we can only conclude $|\Phi| \geq 2$.

**Theorem 5.3** $|\Phi| \geq 5$.

**Proof:** Klazar and Valtr [19] showed that any forbidden subsequence $\sigma$ over three letters has $\text{Ex}(\sigma, n) = O(n)$ unless one of $\sigma_1 = ababa, \sigma_2 = abcacbc, \sigma_3 = abcbcac, \sigma_4 = abacabc, \sigma_5 = abacacb$ is a subsequence of $\sigma$. Notice that $\sigma_1$ is the only palindrome and that $\sigma_2$ and $\sigma_3$ are isomorphic to the reversals of $\sigma_4$ and $\sigma_5$, respectively. Let $\pi_2 = abcaccbc$ and $\pi_8 = abcadcddbd$ be the prototypical forbidden subsequences corresponding to Definition 4.3(2,8); see Figure 8, the second and fifth diagrams on the first row. One can check that $\sigma_2 \prec \pi_2$ but $\sigma_1, \sigma_3, \sigma_4, \sigma_5 \not\prec \pi_2$, and that $\sigma_3 \prec \pi_8$ but $\sigma_1, \sigma_2, \sigma_4, \sigma_5 \not\prec \pi_8$. Thus, either $\pi_2$ (and its reversal) are in $\Phi$ or $\sigma_2$ (and its reversal $\sigma_4$) are in $\Phi$. Similarly, some $\sigma' \prec \pi_8$ (and its reversal) must be in $\Phi$. Notice that all palindrome subsequences of $\pi_8$ are linear, e.g., *abab*, *bccb*, *bdddb*, *bcdcb*. This implies that $\sigma'$ is not isomorphic to its reversal and that it is distinct from $\pi_2, \sigma_2$, and $\sigma_4$. Thus, $\Phi$ contains at least five elements: *ababa* and four distinct subsequences of $\pi_2$ and $\pi_8$ and their reversals. □

# 6 Discussion

It is reasonable to think, for no reasons except those aesthetic, that $\Phi$ is infinite, i.e., that there are infinitely many causes of superlinearity in generalized DS sequences.[2] In this paper we have exhibited the first candidate for $\Phi$, namely $\Psi$, which seems to capture the forbidden structures of the "standard" sequence of $n$ path compressions with length $\Theta(n\alpha(n))$. There are numerous reasons to think that, even if $\Phi$ is infinite, that it does *not* contain $\Psi$ and that it may only share *ababa* in common with $\Psi$. However, as we argue below, some of the obvious objections to the plausibility of $\Psi \subset \Phi$ are not as grounded as one might expect.

One immediate objection is that all $\sigma \in \Psi$ have $\text{Ex}(\sigma, n) = \Omega(n\alpha(n))$. Even if $\Psi$ *did* fully characterize those forbidden subsequences with superlinear growth $\Omega(n\alpha(n))$, why assume that the superlinear spectrum between $\omega(n)$ is $o(n\alpha(n))$ is empty? There's no good response to this objection, except that functions $o(n\alpha(n))$ are exceedingly rare. We are not aware of any natural phenomenon that induces a function $o(n\alpha(n))$, though it is possible to manufacture such functions. Loebl and Nešetřil [22] defined a specialized sequence of path compressions whose length is roughly $n\epsilon(n)$, where $\epsilon$ is the inverse of the quickly growing function corresponding to the ordinal $\epsilon_0$.[3]

---
[2]After $|\Phi| = 1$ is excluded no other cardinality seems quite right.

It is not farfetched to assume that $\text{Ex}(\sigma, n)$ is either $O(n)$ or $\Omega(n\alpha(n))$ for any forbidden subsequence $\sigma$. Even so, the superlinearity of forbidden subsequences in $\Psi$ was established by looking at a *specific* process, namely path compressions, a *specific* sequence of path compressions, and a *specific* way of transcribing path compressions into our ultimate sequence $\Xi$. Is there any reason to believe that $\Xi$ holds a privileged position in the world of generalized DS sequences, despite its apparently ad hoc construction? We think that $\Xi$ is *somewhat* special. This opinion is not grounded in our aesthetic judgement but historical precedent.

Since Tarjan's discovery of the inverse-Ackermann function over 30 years ago we have seen it appear in a wide range of problems. The union-find data structure [25, 28, 30, 11, 20, 13] is certainly the most high profile example. Other examples include one dimensional range searching [33, 5, 9], range searching over trees [27, 29, 26, 8, 31, 24], lower envelopes [32, 2], searching monotone matrices [15, 14], low diameter spanners [7], and, of course, Davenport-Schinzel sequences [12, 2, 18] and related problems concerning forbidden substructure [18]. The fact that the inverse-Ackermann function shows up all over the place is not surprising. (One could rattle off a similar list for any other function.) The surprising part is that all the results above ultimately relate back to one combinatorial object, namely path compressions on balanced trees.[4] Connecting these problems to path compressions is not always direct or simple. To take one example, Klawe's superlinear lower bound on searching monotone matrices [14] is identified with the complexity of the lower envelope of line segments [32], which is identified with ($ababa$)-free Davenport-Schinzel sequences, and generalized postordered path compressions [12] or, equivalently, standard path compressions on balanced trees. In the other examples cited above it is usually easier to convert the domain-specific combinatorial structure, say, a hard instance of one dimensional range searching [9], into a system of path compressions. Given this history it would be truly astounding if it were possible to prove a lower bound of $\text{Ex}(\sigma, n) = \Omega(n\alpha(n))$ using some combinatorial construction that was fundamentally unrelated to path compressions on balanced trees.

Although we see path compressions as the canonical manifestation of the inverse-Ackermann function, we will not argue that $\Xi$ is the only sensible transcription of path compressions. *If* it is possible to prove that, say, $\text{Ex}(abcacbc, n) = \Omega(n\alpha(n))$, the proof would likely use (implicitly or explicitly) the same sequence of path compressions from Section 3 or [25, 12] but a completely different transcription method.

---

[3] Here $\epsilon_0$ is the limit of $\omega, \omega^\omega, \omega^{\omega^\omega}, \ldots$. Although the function corresponding to $\epsilon_0$ shows up in various places, e.g. Goodstein sequences, the Paris-Harrington theorem, and the Hercules-Hydra game, its slowly growing counterpart has yet to appear in a natural way [23, 21].

[4] To paraphrase Tolstoy, all inverse-Ackermann bounds resemble one another, but each linear bound is linear in its own way.